	\documentclass[a4paper,11pt]{article}
	\pdfoutput=1
	\usepackage{jheppub}
	\usepackage[latin1]{inputenc}
	\usepackage{graphicx}
	\usepackage{amsmath}
	\usepackage{amsfonts}
	\usepackage{amssymb}
	\usepackage{mathtools}
	\usepackage{bbold}
	\usepackage{dsfont}
	\usepackage{subfigure}
	\usepackage{tikz}
	\usepackage{xcolor}
	\usetikzlibrary{positioning}
	\usetikzlibrary{arrows}
	\usetikzlibrary{arrows.meta}

	\newcommand{\eqcolon}{\mathrel{\resizebox{\widthof{$\mathord{=}$}}{\height}{ $\!\!=\!\!\resizebox{1.2\width}{0.8\height}{\raisebox{0.23ex}{$\mathop{:}$}}\!\!$ }}}

	\title{\boldmath Radiative Processes of Entangled Detectors in Rotating Frames}

	\author[a]{Gabriel Pican\c{c}o,}
	\author[a]{Nami F. Svaiter,}
	\author[b]{Carlos A. D. Zarro.}

	\affiliation[a]{Centro Brasileiro de Pesquisas F\'{\i}sicas,
	Rua Dr. Xavier Sigaud 150, 22290-180, Rio de Janeiro, RJ, Brazil}
	\affiliation[b]{Instituto de F\'{\i}sica, Universidade Federal do Rio de Janeiro, 21941-972, Rio de Janeiro, RJ, Brazil}

	\emailAdd{gabrielpc@cbpf.br}
	\emailAdd{nfuxsvai@cbpf.br}
	\emailAdd{carlos.zarro@if.ufrj.br}

\abstract{We investigate the radiative processes of accelerated entangled two-level systems. Using first-order perturbation theory, we evaluate transition rates of two entangled Unruh-DeWitt detectors rotating with the same angular velocity interacting with a massive scalar field. Decay processes for arbitrary radius, angular velocities, and energy gaps are analyzed. We discuss the mean-life of entangled states and entanglement harvesting and degradation.}

\date{\today}

\begin{document}

\maketitle
\flushbottom

\section{Introduction}

	Developments in the general theory of quantization of fields in curved spacetime enlarge the possibilities of applications of field theory in our understanding of nature. In canonical quantization, the original construction where quantum states support an irreducible unitary representation of the Poincar\'{e} group must be modified. In this scenario, arbitrary frames for quantization, even in flat spacetime, are laboratories of investigations, since  the vacuum states of quantum fields can be observer-dependent. A quite instructive situation is the quantization performed by uniformly accelerated observers in Minkowski spacetime. The usual treatment for this problem is to quantize a scalar field in the Rindler frame using Rindler's coordinate system \cite{Fulling:1972md}. Both quantizations, in an inertial frame and in a Rindler frame, are unitarily non-equivalents. This can be viewed by analyzing the Bogoliubov's $\beta$ coefficients between Minkowski and Rindler field modes. The fact that the definition of elementary particles and vacuum states for inertial and accelerated observers are distinct can also be viewed by calculating the response function of the detector \cite{Davies75, Unruh:1976db, Birrell:1982ix,Svaiter:1992xt}.  A uniformly accelerated detector interacting with a scalar field prepared in the Poincar\'{e} invariant (Minkowski) vacuum measures a thermal bath, with the temperature being proportional to the proper acceleration. This is known as the Unruh-Davies effect. Moreover, the Unruh-Davies effect anticipates some results of quantum field theories in curved spacetimes, such as the Hawking effect \cite{Hawking:1974sw}.

	The problem of radiative processes of detectors in a non-inertial rotating frame \cite{RotatingBook} can be found in references \cite{Denardo1978, Letaw1980,Bell:1983qr,Bell:1987ir,Doukas:2013noa}. Letaw and Pfausch pointed out that physical content coming from the Bogoliubov's $\beta$ coefficients between the rotating and the inertial modes and the response function of the detector are in disagreement. Since the Bogoliubov's $\beta$ coefficients between the rotating and the inertial modes are zero, the rotating vacuum and the Minkowski vacuum are unitarily equivalent. Nevertheless, the rotating detector interacting with a scalar field in the Minkowski vacuum has a non-zero response function for excitations. This problem was solved by Davies et al \cite{Davies:1996ks}. Besides this incompatibility between response function and Bogoliubov coefficients approaches, other more fundamental problem arises. How to include rotation into a relativistic scenario?

	The answer for this question has attracted many physicists, as for example Ehrenfest, Born, Planck, Kaluza, Einstein, and others \cite{Ehrenfest,Born,Planck,Kaluza,Einstein}. Landau and Lifshitz \cite{landau1975classical} used the transformation law between the cylindrical coordinate system adapted to an inertial frame, and another coordinate system adapted to a rotating one, which is valid only for $r < c/\omega$. In order to extend this coordinate system to any radius, that is, try to solve the problem of tangential velocity being bigger than $c$ for radius $r > c/\omega$, Trocheries and Takeno \cite{Trocheries, Takeno} define a coordinate system adapted to the rotating frame where the tangential velocity is $v/c = \tanh{\omega r/c}$, which only tends asymptotically to $c$. Consequences of this transformation in field theory are discussed in \cite{DeLorenci:1996ym, DeLorenci:2000je, Paola_2001}. This choice is not able to reproduce experimental results, for instance, the Sagnac's effect \cite{Sagnac1,Sagnac2}, where an interferometer in a rotating disk measures the phase shift between two coherent beams of light traveling along path of opposite directions. Another proposal was discussed by Gr{\o}n \cite{Gron,Gron2}. It is able to reproduce Sagnac's effect, but it also has a discontinuity in the time coordinate, for closed circuits around the origin. An alternative approach to discuss the kinematics in rotating frames was developed by Klauber \cite{Klauber}.

	Nowadays, quantum information is a very important topic of research in physics, whether for developing fundamental theory, experiments, or even applications, such as in quantum cryptography or quantum computers \cite{Nielsen:2011:QCQ:1972505}. In particular, relativistic quantum information is becoming always more relevant \cite{Audretsch:1994yz,Peres:2002wx,VerSteeg:2007xs,Hu:2011pd,Martin-Martinez:2014gra,Menezes:2015uaa,Hu:2015lda,Menezes:2015veo,Menezes:2015iva,Menezes:2016quu,Menezes:2017oeb}. The description of detectors coupled to quantum fields claims for the relativistic approach, with measurable effects. One of them is the entanglement degradation \cite{FICEK2002369,breuer2002theory}, where correlated states become uncorrelated by an interaction with a quantum field, for example. This is very important since in realistic experiments we never totally control the coupling of a system to the environment. Another effect is entanglement harvesting \cite{Cliche2011, Martin2012, Salton2015}, where uncorrelated objects become correlated by some other interaction, for example with a quantum field. The interpretation of this phenomenon is that a quantum field in the vacuum state shows correlations between different points in spacetime, and a system coupled with this field can extract entanglement from that. Both effects will be seen in a pair of coupled detectors in a rotating frame, as we will show later.

	In this work, we study a massive scalar field interacting with two Unruh-DeWitt detectors \cite{DeWitt1975}. The detectors are rotating around the origin with the same angular velocity. We discuss radiative processes and quantum entanglement for rotating systems. Using first-order perturbation theory, we calculate the response function of the detectors, looking for the transition rate of excitations or de-excitations between any two arbitrary states. We also compute the mean life of entangled states of the two detectors. Quantum entanglement and quantum harvesting are also discussed in the analysis. We also try to unravel the relevance of all the different parameters in the response function, and consequently we discuss the transition rate.

	This manuscript is organized as follows. In section \ref{sec:quantization} we quantize a massive scalar field in the radially-bounded spacetime, using rotating cylindrical coordinates. In section \ref{sec:detectors} we discuss Unruh-DeWitt detectors. In section \ref{sec:analysis}, we study the response function and derive the expression for the mean life of entangled states. In section \ref{sec:conclusions} we present the conclusions and future directions for this work. In the whole paper, we use $\hbar = c = 1$. The signature of the Minkowski metric in this work is $(+--\,-)$.

\section{Canonical Quantization of a Massive Scalar Field} \label{sec:quantization}

	In this section we discuss the canonical quantization of a massive scalar field in a frame of uniformly rotating observers and also inertial ones. We assume that the coordinates adapted to a rotating, $(t,r,\varphi,z)$, and inertial, $(T,R,\Phi,Z)$ frames are related by the following transformations: 
	\begin{eqnarray}
	T &=& t, \\
	R &=& r, \\
	\Phi &=& \varphi + \omega t, \\
	Z &=& z.
	\end{eqnarray}
	The line element in the rotating frame reads: 
	\begin{equation} \label{eq:rotmetric}
	ds^2 = \left(1 - \omega^2 r^2\right)dt^2 - dr^2 - r^2d\varphi^2 - dz^2 - 2\omega r^2 d\varphi dt.
	\end{equation}
	From the metric (\ref{eq:rotmetric}), we get, as trivial Killing vectors, $\partial_t = (1, 0, 0, 0)$, $\partial_\varphi = (0, 0, 1, 0)$ e $\partial_z = (0, 0, 0, 1)$, the generators of translations on their respective directions. Since the vector $\partial_t$ is not time-like in all spacetime, we cannot define positive and negative modes for all radial coordinate. This definition will be discussed latter. This problem was solved imposing Dirichlet's boundary conditions for $r = \omega^{-1}$ \cite{Davies:1996ks}. Another Killing vector, time-like in all spacetime, $\partial_T = (1, 0, -\omega, 0)$, which is the generator of translation in the time coordinate adapted to inertial frames, will be useful in our discussions. 

	 In order to implement the canonical quantization, we have to solve the Klein-Gordon equation in the rotating frame \cite{Denardo1978,Letaw1980}:
	\begin{equation}\label{eq:KG}
	\left(\partial^2_t  - \frac{1}{r} \partial_r (r \partial_r ) - \left( \frac{1}{r^2} - \omega^2 \right) \partial^2_\varphi  - \partial^2_z  - 2\omega \partial_t \partial_\varphi  + \mu^2\right) \phi = 0,
	\end{equation}
	where $\mu$ is the mass of the scalar field. To proceed, let us make an ans\"{a}tz for the complete set of modes $u_{\varepsilon m k}$
	\begin{equation}\label{eq:RotMode}
	u_{\varepsilon m k}(t, r, \varphi, z) \propto \exp \bigl( -i\varepsilon t + im\varphi +ikz \bigr) R(r),
	\end{equation}
	where $\varepsilon$, $m$, and $k$ are arbitrary constants that label the field modes. Substituting equation  (\ref{eq:RotMode}) into equation (\ref{eq:KG}), we obtain the radial equation
	\begin{equation}
	\frac{1}{r} \frac{d}{dr} \left( r \frac{dR(r)}{dr} \right) + \left( (\varepsilon + m\omega)^2 - k^2 - \mu^2 - \frac{m^2}{r^2} \right) R(r) = 0.
	\end{equation}
	The physical acceptable solutions for the above equation are Bessel functions of first kind, $J_m$. Defining $(\varepsilon + m\omega)^2 - k^2 - \mu^2 = \chi^2$, the radial solution can be written as
	\begin{equation}
	J_m (\chi r) = \sum_{n=0}^\infty \frac{(-1)^n}{n! \Gamma (n+m+1)} \left( \frac{\chi r}{2} \right)^{m+2n}, \quad n \: \epsilon \: \mathbb{Z}.
	\end{equation}
	The Dirichlet's boundary conditions on the radial coordinate is 
	\begin{equation}
	J_m(\chi a) = 0.
	\end{equation}
	Therefore, $\chi = \alpha_{mn}/a = k_{mn}$, where $\alpha_{mn}$ is the $n$-th root of the $m$-th Bessel function of first kind. In this case, the normalization of the radial mode is given by:
	\begin{equation}
	\int_0^a dr \: r J_m(k_{mn} r) J_m(k_{ml} r) = \frac{a^2}{2} [J'_m(k_{mn} a)]^2 \delta_{nl},
	\end{equation}
	where $J'_m(k_{mn} a)=\left.\frac{d J_m(k_{mn} r)}{dr}\right|_{r=a}$. The normalized cylindrical modes in the rotating frame are written as
	\begin{equation} \label{eq:NormRotModes}
	u_{kmn}(t,r,\varphi,z) = \frac{\exp[-i\varepsilon t +im\varphi +ikz] J_m(k_{mn} r)}{2\pi a [J'_m(k_{mn} a)] N_{kmn}},
	\end{equation}
	where $N_{kmn}$ refers to the different possible normalization given by the two time-like Killing vectors. It reads
	\begin{equation} \label{eq:normalization} 
	N_{kmn} = \sqrt{\varepsilon},\, \text{if}\, K^\mu = \partial^{\mu}_{T}, \quad \text{and} \quad N_{kmn} = \sqrt{\varepsilon + m\omega},\, \text{if}\, K^{\mu} = \partial^{\mu}_{t}.
	\end{equation}

	In this scenario, it is natural to define an inner product for each of the possible time-like Killing vectors $K^\mu$, between two arbitrary field modes $\psi_{i}$ and $\psi_{j}$, where $i$ and $j$ are arbitrary indexes labelling the modes, in the following way:
	\begin{equation}
	\langle \psi_{i}, \psi_{j} \rangle = i\int \sqrt{|h|} d\Sigma_\mu \left[\psi_{i}^{*} \,K^\mu\, \psi_{j} - \psi_{j} \,K^\mu\, \psi_{i}^{*}\right],
	\end{equation}
	where $d\Sigma_\mu$ is the future-oriented volume element of $\Sigma$ and $h$ is the determinant of the metric induced in the hypersurface. Since the inner products between arbitrary field modes are
	\begin{equation}
	\langle u_{kmn}, u_{k'm'n'}^{*}\rangle \: = \: \langle u_{kmn}^* , u_{k'm'n'}\rangle \: =\: 0
	\end{equation}
	and
	\begin{equation}
	\langle u_{kmn} , u_{k'm'n'}\rangle\: =\: - \langle u_{kmn}^* , u_{k'm'n'}^*\rangle\: =\: \delta(k-k') \delta_{m m'} \delta_{n n'},
	\end{equation}
	we say that $u_{kmn}$ and $u^*_{kmn}$ are positive and negative norm modes, respectively. Notice that they are also respectively positive and negative frequency modes with respect to the time coordinate adapted to the rotating frame. Introducing $E$ such that $E^2 := (\varepsilon + m\omega)^2 = k_{mn}^2 + k^2 + \mu^2$, we get 
	\begin{equation}
	-i\varepsilon t + im\varphi = -i(E-m\omega)t + im\varphi = -iEt + im(\varphi + \omega t) = -iET + im\Phi. 
	\end{equation}

	We shall be concerned with the canonical quantization of the scalar field in the rotating frame. Defining $u_i (x)$ and $u_i^* (x)$ as the field modes and their complex conjugates in the rotating frame, one can expand the massive scalar field in the form
	\begin{equation}
	\phi(x) = \sum_i \left[a_i u_i(x) + a_i^\dagger u_i^*(x) \right],
	\end{equation}
	where $i$ is a generic set of index and $a_i$ and $a_i^\dagger$ are respectively the annihilation and the creation operators associated to field modes. This expansion defines the rotating vacuum.

	In order to compare both quantizations using the inertial modes and the rotating modes, one can compute the Bogoliubov coefficients between these modes. One shows that the Bogoliubov's $\beta$ coefficients are zero, since
	\begin{equation}
	u_{kmn} (t, r, \varphi, z) \propto \exp[-i\varepsilon t +im\varphi + ikz]J_m(k_{mn} r)
	\end{equation}
	and
	\begin{equation}
	U_{kmn} (T, R, \Phi, Z) \propto \exp[-iET +im\Phi + ikZ]J_m(k_{mn} R),
	\end{equation}
	where $U_{kmn}$ are positive frequency modes with respect to the time of the inertial frame, in cylindrical coordinates. Therefore, the vacuum expectation value of one frame's number operator calculated in the other frame's vacuum state is always zero. In the following, it is important to define the positive Wightman function $G_{jk}^+ (x_j, x_k) = \langle 0_R|\phi(x_j) \phi(x_k)|0_R \rangle$, which can be obtained using the positive field modes  (\ref{eq:NormRotModes}) as
	\begin{equation}\label{pwf}
	G_{jk}^+(x_j, x_k) = \sum_{m=-\infty}^\infty \sum_{n=1}^\infty \int_{-\infty}^\infty dk \frac{J_m(k_{mn}r_j) J_m(k_{mn}r_k) e^{-i [\varepsilon \Delta t -m \Delta \varphi - k\Delta z] }}{4 \pi^2 a^2 [J'_m(k_{mn}a)]^2 N_{kmn}^2},
	\end{equation}
	where $\Delta x^\mu = x^\mu_j - x^\mu_k$.

\section{Radiative Processes of non-Inertial Entangled Detectors} \label{sec:detectors}

\begin{figure}
	\centering 
	\begin{tikzpicture}
	\draw[very thick] (3,1) -- (6,1);
	\draw[very thick] (3,7) -- (6,7);
	\draw[very thick] (0,4) -- (3,4);
	\draw[very thick] (6,4) -- (9,4);
	\draw[dashed] (3.2,4) -- (5.8,4);
	\draw[{Latex[length=2.5mm,width=2.5mm]}-{Latex[length=2.5mm,width=2.5mm]}] (4,7)--(1.5,4);
	\draw[{Latex[length=2.5mm,width=2.5mm]}-{Latex[length=2.5mm,width=2.5mm]}] (5,7)--(7.5,4);
	\draw[{Latex[length=2.5mm,width=2.5mm]}-{Latex[length=2.5mm,width=2.5mm]}] (4,1)--(1.5,4);
	\draw[{Latex[length=2.5mm,width=2.5mm]}-{Latex[length=2.5mm,width=2.5mm]}] (5,1)--(7.5,4);
	\node at (2.5, 1) {$|g\rangle$};
	\node at (-0.5, 4) {$|a\rangle$};
	\node at (9.5, 4) {$|s\rangle$};
	\node at (2.5, 7) {$|e\rangle$};
	\node at (5, 2.5) {$\frac{E}{2}$};
	\draw[<->, gray] (4.5,1)--(4.5,4);
	\node at (5, 5.5) {$\frac{E}{2}$};
	\draw[<->, gray] (4.5,4)--(4.5,7);
	\draw[->] (-1.5, 0.5) -- (-1.5, 7.5);
	\draw[] (-1.7, 4) -- (-1.3, 4);
	\node at (-2, 4) {0};
	\node at (-2, 7.5) {E};
	\end{tikzpicture}
	\caption{Energy levels and possible transitions between the eigenstates of the detectors' Hamiltonian (\ref{eq:Hd}), with $\Omega$ = 0. Adapted from Ficek et al \cite{FICEK2002369}.}
	\label{fig:eigenstates}
	\end{figure}
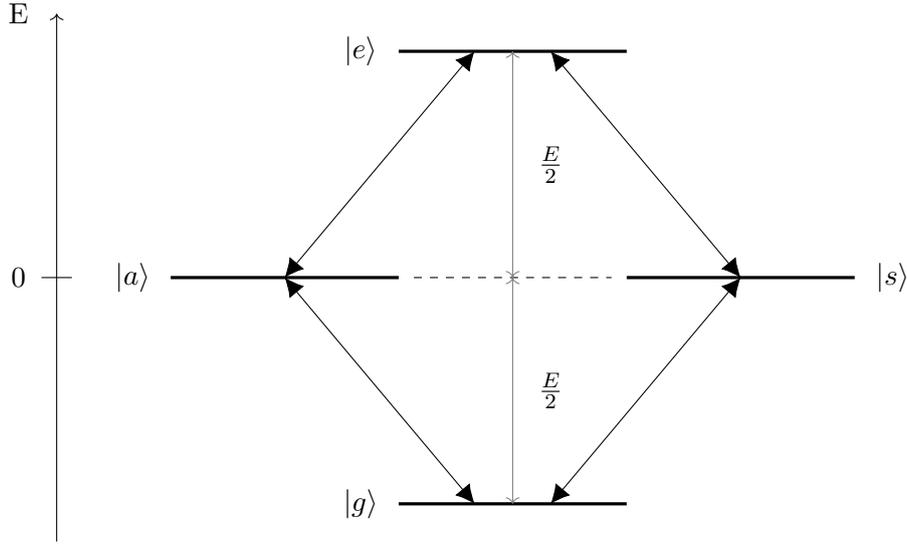

	The aim of this section is to discuss radiative processes of non-inertial entangled detectors. We assume two identical Unruh-DeWitt detectors coupled to a massive scalar field (see also \cite{MartinMartinez:2012th, Alhambra:2013uja, Martin-Martinez:2015psa, Pozas-Kerstjens:2016rsh,Arias:2015moa}).	The total Hamiltonian of the system is given
	\begin{equation}
	H = H_d + H_f + H_{int},
	\end{equation}
	where $H_d$ and $H_f$ are the free detectors and field Hamiltonians, respectively. The $H_{int}$ is the interaction Hamiltonian between the two-level systems and the scalar field. With $|g_j \rangle$ and $|e_j \rangle$ being respectively the ground state and the excited state of the $j$-th detector, the free Hamiltonian of the two detectors in their proper time is given by
	\begin{equation} \label{eq:Hd}
	H_d = \frac{E}{2} \left[ S^z_1\otimes \mathds{1}_2 + \mathds{1}_1 \otimes S^z_2 \right] + \Omega(S^+_1 S^-_2 + S^-_1 S^+_2),
	\end{equation}
	where $S^z_j = |e_j\rangle\langle e_j| - |g_j\rangle\langle g_j|$, $S^+_j = |e_j\rangle\langle g_j|$ and $S^-_j = |g_j\rangle\langle e_j|$, for j = 1, 2. The detector Hamiltonian can be diagonalized, we obtain the following four orthogonal bases states:
	\begin{align}
	|g\rangle &= |g_1\rangle |g_2\rangle;\\    
	|a\rangle &= \frac{1}{\sqrt{2}} \left( |g_1\rangle|e_2\rangle - |e_1\rangle|g_2\rangle \right);\label{eq:a}\\
	|s\rangle &= \frac{1}{\sqrt{2}} \left( |g_1\rangle|e_2\rangle + |e_1\rangle|g_2\rangle \right)\label{eq:s};\\
	|e\rangle &= |e_1\rangle|e_2\rangle,
	\end{align}
	with eigenvalues $-E$, $-\Omega$, $+\Omega$ and $+E$, respectively. A tensor product is implicit in the above notation. The states $|a\rangle$ and $|s\rangle$ are maximally entangled states. The ground state of both detectors is $|g\rangle$, $|s\rangle$ and $|a\rangle$ are the symmetric and anti-symmetric states, respectively, and $|e\rangle$ is the state where both detectors are excited. Now, the free Hamiltonian of the massive scalar field $\phi$ is given by
	\begin{equation}
	H_{f} = \frac{1}{2} \int d^3 x \left[ \bigl(\dot{\phi}(x)\bigr)^2 + \bigl(\nabla \phi(x)\bigr)^2 + \mu^2 \phi^2(x) \right],
	\end{equation}
	where $\mu$ is the mass of the field, the dot represents derivative with respect to $t$ and $\nabla$ is the gradient operator.  Finally  the interaction Hamiltonian is written as
	\begin{equation}
	H_{int}(t) = \lambda \sum_{j=1}^{2} \chi_j\bigl(\tau_j(t)\bigr)m^{(j)}\bigl(\tau_j(t)\bigr) \phi\bigl(x^\mu\left(\tau_j(t)\right)\bigr)\frac{d\tau_j(t)}{dt},
	\end{equation}
	where $\lambda$ is the dimensionless coupling constant of the interaction, $\chi$ is a real-valued switch-function for the interaction of the detectors with the scalar field, and $m^{(j)}(\tau_j(t))$ is the monopole operator of the $j$-th detector. The field $\phi(x^\mu(\tau_j))$ is evaluated in the classical trajectory of each of the detectors, and the factor $d\tau_j/dt$ is the Jacobian to correct the time integration. 
	The operators, $m^{(1)}(0)$ and $m^{(2)}(0)$, for two detectors in the basis $\{|g\rangle, |s\rangle, |a\rangle, |e\rangle\}$ are 
	\begin{equation}
	m^{(1)}(0) = m(0) \otimes \mathds{1} = \frac{1}{\sqrt{2}} \begin{bmatrix} 0 & m & m & 0 \\ m & 0 & 0 & m \\ m & 0 & 0 & -m \\ 0 & m & -m & 0 \end{bmatrix}, \label{eq:m1}
	\end{equation}
	\begin{equation}
	m^{(2)}(0) = \mathds{1} \otimes m(0) = \frac{1}{\sqrt{2}} \begin{bmatrix} 0 & m & -m & 0 \\ m & 0 & 0 & m \\ -m & 0 & 0 & m \\ 0 & m & m & 0 \end{bmatrix}. \label{eq:m2}
	\end{equation}
	In the interaction picture, for arbitrary initial and final states $|i\rangle$ and $|f\rangle$ of the detectors, respectively, we have
	\begin{equation}
	\langle f|m^{(j)}(\tau_j)|i\rangle = e^{i(E_{f} - E_{i}) \tau_{j}}\langle f|m^{(j)}(0)|i\rangle = e^{i(E_f - E_i) \tau_j} m^{(j)}_{fi}.
	\end{equation}
	where $E_i$ and $E_f$ are the energies of the initial and final detector states, respectively. The only possible transitions are the ones shown in figure \ref{fig:eigenstates}, where both $m^{(j)}_{fi} \neq 0$. For simplicity, we take $\Omega = 0$, such that the Bell states are degenerated. The energy levels are also illustrated in figure \ref{fig:eigenstates}.

	To calculate the probability of transition between arbitrary states, we use  the Schr\"{o}dinger equation of the interaction picture
	\begin{equation}
	i\frac{dU(t, t_i)}{dt} = H_{int}(t)U(t, t_i),
	\end{equation}
	such that
	\begin{equation}
	U(t, t_i) = \mathcal{T} \left\{ \exp \left( -i \int_{t_i}^t H_{int}(t')dt' \right) \right\} = 1 -i\lambda \int_{t_i}^{t_f} dt H_{int}(t) + \mathcal{O}(\lambda^2),
	\end{equation}
	where $t_i$ is an arbitrary initial time, and $\mathcal{T}$ is the usual time-ordering operator. With the evolution operator, one can compute the transition amplitude between arbitrary states $|t_i\rangle=|i\rangle \otimes |\phi_{i}\rangle$ and $|t_f\rangle=|f\rangle \otimes |\phi_{f}\rangle$. We get
	\begin{equation}
	A_{|t_i\rangle \to |t_{f}\rangle} = (\langle f| \otimes \langle\phi_f|)U(t_f, t_i) (|i\rangle \otimes |\phi_i\rangle),
	\end{equation}
	where $|\phi_i\rangle$ and $|\phi_f\rangle$ are the initial and final states of the scalar field. Assuming the initial state of the field as the rotating vacuum state, $|\phi_i\rangle = |0_R\rangle$, and tracing out $|\phi_f\rangle$, we get that the probability of transition can be written as
	\begin{align}
	P_{|0_R, i\rangle\rightarrow |f\rangle} = & \lambda^2 \int_{t_i}^{t_f} dt \int_{t_i}^{t_f} dt' \sum_{j=1}^2 \sum_{k=1}^2 m_{fi}^{(j)*} m_{fi}^{(k)} \langle 0_R|\phi\bigl(x^\mu_j(\tau_j(t))\bigr)\phi\bigl(x'^\mu_k(\tau_k(t'))\bigr)|0_R\rangle \nonumber \\
	&\times \frac{d\tau_j(t)}{dt} \frac{d\tau_k(t')}{dt'} e^{-i(E_f - E_i)(\tau_j-\tau_k)} \chi(\tau_j) \chi(\tau_k),
	\end{align}
	where $j$ and $k$ label both detectors.	This probability of transition is a combination of products of two factors: the selectivity $m_{fi}^{(j)*} m_{fi}^{(k)}$, only involving detectors' internal structure, and the response function $F_{jk}$, describing the interaction with the field, as follows:
	\begin{align}
	F_{jk}(\Delta E, \chi, t_i, t_f) &=  \int_{t_i}^{t_f} dt \int_{t_i}^{t_f} dt'  G^+_{jk}(t, t') \frac{d\tau_j(t)}{dt} \frac{d\tau_k(t')}{dt'}  e^{-i\Delta E(\tau_j-\tau_k)} \chi(\tau_j) \chi(\tau_k),
	\end{align}
	where  $G_{jk}^+ (x_j, x_k') = \langle 0_R|\phi(x_j) \phi(x_k')|0_R\rangle$ is the positive Wightman function associated to the scalar field, as discussed in equation (\ref{pwf}). With the above definitions the probability of transition between two arbitrary states is given by
	\begin{equation}
	P_{|0_R, i\rangle\rightarrow |f\rangle} = \lambda^2 \sum_{j,k=1}^2 m_{fi}^{(j)*} m_{fi}^{(k)} F_{jk} (\Delta E, \chi, t_i, t_f).
	\end{equation}

	From now, we will generalize the results obtained by Cai, Li and Ren \cite{Cai:2018xvz}. See also reference \cite{Doukas:2013noa} for a rotating Unruh-DeWitt detector under non-equilibrium conditions and reference \cite{Rodriguez-Camargo:2016fbq} for a discussion of a finite-time response function. Let us use equation (\ref{pwf}) into the response function to calculate transition rates of a system of two Unruh-DeWitt detectors in a uniformly rotating frame, with different radial coordinates. Without loss of generality, to simplify our computations, we can use $z_1 = z_2$ and $\varphi_1 = \varphi_2$. We get the proper times $\tau_j = \left({1-w^2r_j^2}\right)^{1/2} t = t /\gamma_j$. The term $d \tau_j / dt$ is constant in a circular motion, and can be factored out of time integrals. The response function reads:
	\begin{equation}\label{eq:Fantes}
	F_{jk} = \int_{t_i}^{t_f} dt \int_{t_i}^{t_f} dt' \sum_{m=-\infty}^\infty \sum_{n=1}^\infty \int_{-\infty}^\infty dk \frac{J_m(k_{mn}r_j) J_m(k_{mn}r_k) e^{-i[\Delta E (\tau_j - \tau_k') + \varepsilon (t-t')]}}{4\pi^2 a^2 \gamma_j \gamma_k [J'_m(k_{mn}a)]^2 N_{kmn}^2}.
	\end{equation}
	Let us change variables of integration from $t$ and $t'$ to $t$ and $\Delta t = t - t'$. The modulus of the Jacobian for this coordinate transformation is one. Let us define $\Delta \bar{E} = \Delta E(1/\gamma_j - 1/\gamma_k)$ and $\Delta E' = \Delta E / \gamma_k$, and rewrite the exponential argument as follows:
	\begin{equation}\label{eq:ChangeVar}
	\exp \left[ -i\varepsilon (t - t') -i\Delta E \left(\frac{t}{\gamma_j} - \frac{t'}{\gamma_k}\right) \right] = \exp \left[ -i(\varepsilon + \Delta E')(t - t') -i\Delta \bar{E}t \right].
	\end{equation}
	We will work with the asymptotic limits $t_i \rightarrow -\infty$ and $t_f \rightarrow \infty$. 

	Now, the response function per unit time $t$, the rate $R_{jk}(t) = \partial F_{jk}/\partial t$ can be computed
	\begin{align}\label{eq:Fdepois}
	R_{jk} = e^{-i\Delta \bar{E}t} \sum_{m, n} \int_{-\infty}^\infty dk \frac{J_m(k_{mn}r_j) J_m(k_{mn}r_k)}{4\pi^2 a^2 \gamma_j \gamma_k [J'_m(k_{mn}a)]^2 N_{kmn}^2}& \int_{-\infty}^\infty d(\Delta t) e^{-i (\Delta t) (\varepsilon + \Delta E')} \nonumber \\
	 = e^{-i\Delta \bar{E}t} \sum_{m, n} \int_{-\infty}^\infty dk \frac{J_m(k_{mn}r_j) J_m(k_{mn}r_k)}{4\pi^2 a^2 \gamma_j \gamma_k [J'_m(k_{mn}a)]^2 N_{kmn}^2}& \left [ 2\pi \delta (\varepsilon + \Delta E') \right ],
	\end{align}
	where the delta function on the last equality was obtained by performing the integral on $\Delta t$, resulting in a factor $\delta[\Delta E' + (\sqrt{k^2 + k^2_{mn} + \mu^2} - m\omega)]$ after substituting $\varepsilon$. 

	The roots of the Bessel function are such that $\alpha_{mn} > m$, so, as $\omega a \leq 1$, the argument of the delta function is always positive, and the corresponding response function will be zero \cite{Davies:1996ks}. We will have non-zero response function and non-zero contribution to the transition rate if and only if $\Delta E' \leq 0 \leq m\omega - \sqrt{k_{mn}^2 + \mu^2}$. This means that there is no excitation of a inertial detector in the rotating vacuum, it can only de-excite. It is consistent with the fact that all Bogoliubov's $\beta$ coefficients are zero between the rotating and inertial modes. So inertial detectors can not detect particles in the rotating vacuum.

	Assuming that $\Delta E' \leq m\omega - \sqrt{k_{mn}^2 + \mu^2}$, we can expand the delta function in its roots, $\delta\bigl(f(k)\bigr) = \sum_{i-th \hspace{0.1cm} root} \left(\frac{\delta(k - k_i)}{|f'(k_i)|}\right)$, with
	\begin{equation}
	f'(k) = \frac{k}{\sqrt{k^2 + k^2_{mn} + \mu^2}} = \frac{\sqrt{(m\omega - \Delta E')^2 - k^2_{mn} - \mu^2}}{m\omega - \Delta E'}.
	\end{equation}
	Integrating the expanded delta function, the response function becomes:
	\begin{equation}\label{eq:RespRateCjk}
	\hspace{-0.35cm} R_{jk} = \frac{e^{-i\Delta \bar{E}t}}{2 \pi a^2 \gamma_j \gamma_k} \sum_{m, n} \frac{J_m(k_{mn} r_j) J_m(k_{mn} r_k) |m\omega - \Delta E'|}{[J_{m+1}(k_{mn} a)]^2 N_{kmn}^2 \sqrt{|(m \omega - \Delta E')^2 - k^2_{mn} - \mu^2|}} \eqcolon e^{-i\Delta \bar{E}t} C_{jk},
	\end{equation}
	%
	where we defined the numerical factor $C_{jk}$ as all the terms in the above equation that does not depend on the time $t$. In the following, we show that $R_{jk}$ is real, as expected. This will be discussed later.

	We can express, in first-order perturbation theory, the transition rate $\dot{P} =  dP/dt$ as
	\begin{equation}
	\Gamma_{|i\rangle \rightarrow |f\rangle} = \dot{P}_{|i\rangle \rightarrow |f\rangle} = \lambda^2 \sum_{j,k=1}^2 m_{fi}^{(j)*} m_{fi}^{(k)} R_{jk}. \label{eq:transrate}
	\end{equation}
	If we compute the $C_{jk}$ numerical factor, we are able to study the allowed radiative processes in this system, and its transition rates. But this factor is not fully determined yet in (\ref{eq:RespRateCjk}), we still need to specify the normalization used. We will compute it using both possible normalizations $N_{kmn}$ ($\ref{eq:normalization}$) discussed in the previous section, for which the factor $C_{jk}$ reads:
	\begin{align}
	K &= \partial_T:  \nonumber \\
	& C_{T_{jk}} = \frac{1}{\gamma_j \gamma_k}\sum_{m,n} 
	\frac{J_m(k_{mn} r_j) J_m(k_{mn} r_k) \Theta[m\omega - \sqrt{k_{mn}^2 + \mu^2} - \Delta E']}{2 \pi a^2 [J_{m+1}(k_{mn} a)]^2 \sqrt{(m \omega - \Delta E')^2 - k^2_{mn} - \mu^2}}. \label{eq:CT}\\
	K &= \partial_t: \nonumber \\
	& C_{t_{jk}} = \frac{1}{\gamma_j \gamma_k}\sum_{m,n}
	\frac{J_m(k_{mn} r_j) J_m(k_{mn} r_k) |m\omega - \Delta E'|\Theta[m\omega - \sqrt{k_{mn}^2 + \mu^2} - \Delta E']}{2 \pi a^2 [J_{m+1}(k_{mn} a)]^2 (-\Delta E') \sqrt{(m \omega - \Delta E')^2 - k^2_{mn} - \mu^2}}. \label{eq:Ct}
	\end{align}

\section{Analysis of the Radiative Processes}\label{sec:analysis}

	\subsection{Discussion of the Response Function}\label{subsec:discussion}

	The response function presented in equations (\ref{eq:CT}) and (\ref{eq:Ct}) is a product of the integral of an oscillatory term in $t$ with a numerical factor called $C_{jk}$. Defining the rate $R_{jk}(t)$ as usual, being the derivative $dF_{jk}(t)/dt$ of the response function, the first term becomes only a phase. Notice that the phase $\Delta \bar{E} \, t = 0$ for both $R_{11}$ and $R_{22}$, such that these terms will never become negative when calculating the transition rate $\Gamma$, as $\Delta \bar{E} = \Delta E (1/\gamma_j - 1/\gamma_k)$. The phase of the crossed terms, $R_{12}$ and $R_{21}$, are complex conjugates, so, when summed, they will only result in a real factor times a trivial oscillatory term. In fact, there is no origin defined for the time coordinate, so we can specify it stating that we are performing the calculations to $t=0$, which is equivalent to taking the absolute value of each $R_{jk}$.

	It is worth to note that, when we changed variables in equation (\ref{eq:ChangeVar}), we chose the time $t$ of the first detector as the variable for the response function. We could have chosen the time $t'$ of the second detector, and the only effect would be that $\Delta E'$ would be equal to $\Delta E / \gamma_j$ instead of $\Delta E / \gamma_k$. That is, the choice of the time coordinate to describe the system only implies in which gamma factor will Doppler shift the gap of the detector in our description of the system.

	We have two possible Killing vectors defining our internal product, $\partial_T$ and $\partial_t$, being the generator of temporal displacements in the non-rotating and in the rotating frames, respectively. Using suitable boundary conditions, both are time-like in all of the radially-bounded spacetime. The difference between those normalizations is given by a term $|m\omega - \Delta E'|/|\Delta E'|$ = $| 1 - m\omega/\Delta E' |$ inside the sums. It can only be significant for $m\omega \approx \Delta E'$. According to the convergence criterion described in the next subsection, we always had $m_{max} \leq 200$, so we need $\omega \approx 0.1$ for this term to be relevant. We will call it a ``non-relativistic regime" when $\omega \ll 0.1$, and a ``relativistic regime" otherwise. In fact, when we compare the numerical results for the transition rates, we confirm the values of the transition rates with both normalizations begin to differ only in the relativistic regime, but none of the qualitative features will differ between them (cf. figure \ref{fig:CTandCt}).

	Despite having eight possible transitions shown in figure \ref{fig:eigenstates}, we will find that we have only three different transition rates. The first one is related to de-excitations involving the symmetric entangled state $\left(\Gamma_{|e\rangle \rightarrow |s\rangle}= \Gamma_{|s\rangle \rightarrow |g\rangle}\right)$; the second, to de-excitations involving the anti-symmetric entangled state $\left(\Gamma_{|e\rangle \rightarrow |a\rangle}=\Gamma_{|a\rangle \rightarrow |g\rangle}\rangle\right)$, and, lastly, the third one, involving any excitation $\left(\Gamma_{|g\rangle \rightarrow |s\rangle}=\Gamma_{|s\rangle \rightarrow |e\rangle}=\Gamma_{|g\rangle \rightarrow |a\rangle}=\Gamma_{|a\rangle \rightarrow |e\rangle}\right)$. As any of the possible transitions necessarily involve one pure state and one entangled state, any of them by themselves represent either entanglement degradation or entanglement harvesting. 

	In order to compute transition rates, we have to combine the rates (individual rates $R_{11}$ and $R_{22}$ and crossed rates $R_{12}$ and $R_{21}$) with the selectivity factors. Both kinds of response functions have sums of products of Bessel cylindrical functions, which have strong oscillatory behavior. The individual rates show products of those functions taken at the same point, so, as they are squared, these terms will never be negative. Only the crossed rates can be negative. Therefore, when we take both detectors to the same radial coordinate, $r_1 = r_2$, the crossed response functions will also be necessarily positive. It is expected that in this situation we would get at least a local maximum in the transition rate, for any equal radial coordinates. We found it to be evidently a global maximum in all explicitly calculated cases, and one of them is exhibited in the next subsection. This behavior has been widely discussed in the literature \cite{Ford:1994zz}. When the detectors are too close, they interfere stronger with each other. In fact, as we will see (cf. figure \ref{fig:C's}), the crossed response functions is only significantly different from zero when $r_1 \approx r_2$.

	Due to the dependence $ 1/\sqrt{(m\omega - \Delta E')^2 - k_{mn}^2 - \mu^2} $ in the response function, divergences can appear. In the next subsection, we see them clearly as peaks in the plot of the transition rate by $\omega a$ when it approaches one, as we see in figure \ref{fig:RelatSym}(c), with $|\Delta E|a = 200$. Looking into the denominator of $C_{jk}$, we see that these singularities only happen when $m\omega \approx \Delta E' \rightarrow ma\omega \approx |\Delta E'|a$, or when $ k_{mn} \approx \Delta E' \rightarrow \alpha_{mn} \approx |\Delta E'|a $. Since in the sums $m_{max}$ and $n_{max} < 200$, let us take $\alpha_{200,200}$ as superior limit for $\alpha_{mn}$. We have $\alpha_{200,200} \approx 920 > 200$. So, for $|\Delta E'|a = 200$, we have $m$ and $n$ such that $\alpha_{mn} \approx |\Delta E'|a \approx ma\omega$. In this case, we will have singularities from both terms, in many of $(m,n)$ pairs. If we had chosen $a$ one order of magnitude bigger, we would not expect any singularity. If we fix $a$, but reduce $\omega$, we will not have divergences caused by the factor $m\omega$, but we may still have some singularities coming from $k_{mn}$. Since the only place where $\mu$ appears is in this factor, we can say that its main effect is to change the regime when we start having singularities.

	For $\gamma_1, \gamma_2 \approx 1$, we can go back into equation (\ref{eq:CT}) and take the approximation $\Delta E' \approx \Delta E$, such that the only dependence on the radial coordinates will be in the Bessel functions. We will specify the details of this approximation for the normalization using the $\partial_T$ Killing vector, but for $\partial_t$ it would be basically the same, just including the factor $|1 - m\omega/\Delta E|$ in the normalization. We approximate the $C_T$-factor as

	\begin{equation}
	    C_{T_{jk}} \approx \sum_{m,n} T_{mn} J^{mn}_j J^{mn}_k,
	\end{equation}
	where
	\begin{equation}
	    T_{mn} = \frac{\Theta[m\omega - \sqrt{k_{mn}^2 + \mu^2} - \Delta E]}{2 \pi a^2 [J_{m+1}(k_{mn} a)]^2 \sqrt{(m \omega - \Delta E)^2 - k^2_{mn} - \mu^2}}
	\end{equation}
	and
	\begin{equation}
	    J^{mn}_j = \frac{J_m(k_{mn} r_j)}{\gamma_j}.
	\end{equation}
	Now, the rate can be written in a much simpler way. Using the matrix elements of (\ref{eq:m1}) and (\ref{eq:m2}) in equation (\ref{eq:transrate}), we get two main cases, transitions involving the symmetric entangled state, and transitions involving the anti-symmetric entangled state:
	\begin{equation}
	    \Gamma'_{symm} = \lambda^2 \sum_{m,n} T_{mn} (J^{mn}_1 + J^{mn}_2)^2; \quad \Gamma'_{anti-symm} = \lambda^2 \sum_{m,n} T_{mn} (J^{mn}_1 - J^{mn}_2)^2, \label{eq:SandA}
	\end{equation}
	where the $\Gamma'$ represents an approximated transition rate. It is clear, from these equations, that transition rates involving the anti-symmetric entangled state are zero when $r_1 = r_2$. Moreover, only for $r_1 = r_2$ we know that $J^{mn}_1$ and $J^{mn}_2$ have the same signal for all $m$'s and $n$'s. So, it can also be expected that this point is the maximum of transition rates involving the symmetric entangled state. In the next sub-section, we compare numerical analysis using the approximated transition rates in equation (\ref{eq:SandA}), and the ones calculated using the functions in equation (\ref{eq:CT}).

	\subsection{Numerical analysis of Radiative Processes}\label{subsec:numerical}

	This section is devoted to the study of numerical values for the transition rates of some interesting cases, revealing the behavior of the system. Our convergence criterion for the sums in $m$ and $n$ was that the relative difference between the following terms of the sum should be less than $10^{-7}$, 10 times in a row. Although there are sums that do not converge for $m, n \leq 100$, for $m, n \leq 200$ all of them converged. All of the following plots have dimensionless quantities in both axes. Unless we explicitly say otherwise, the default values for the parameters are such that $|\Delta E|a = 20000$, $a\omega = 1$, $\mu/|\Delta E| = 0.035$ and $\Delta E = -20$ (in arbitrary units of energy). For simplicity, we also took $\lambda = 1$, and the monopole operator constant $m^{(1)} = m^{(2)} = \sqrt{2}$.

\begin{figure}[h!]
	    \centering
	     \subfigure[ref1][$C_{11}(r_2)$]{\includegraphics[width=.45\linewidth]{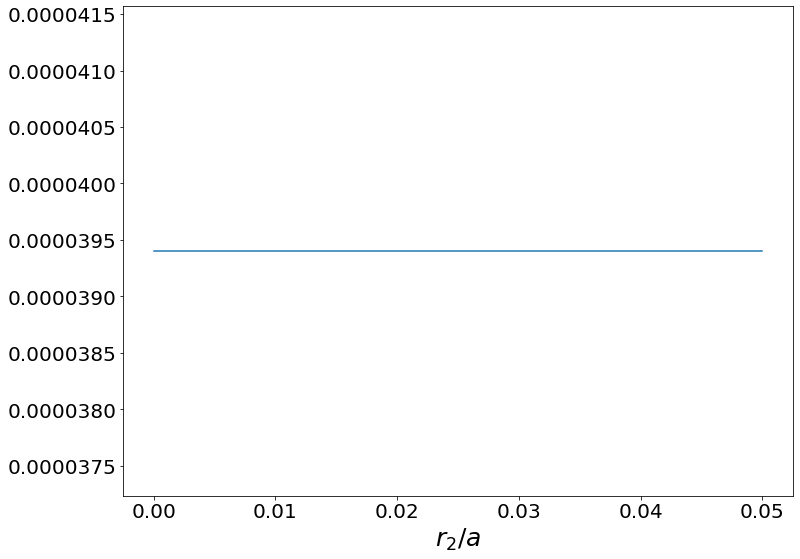}}\qquad
	     \subfigure[ref1][$C_{22}(r_2)$]{\includegraphics[width=.45\linewidth]{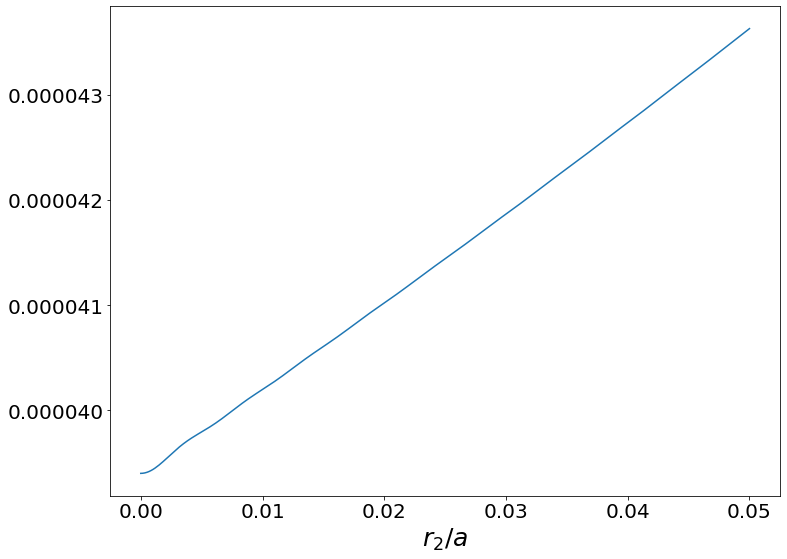}}\\
	     \subfigure[ref1][$C_{12}(r_2)$]{\includegraphics[width=.45\linewidth]{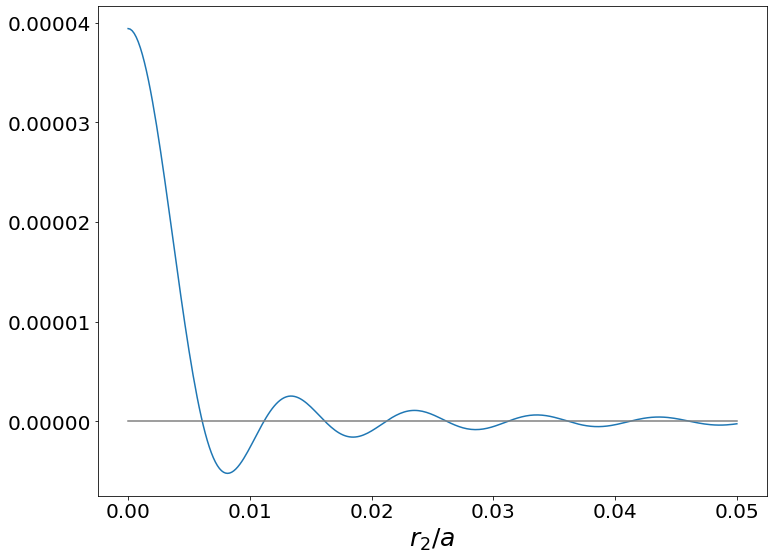}}\qquad
	     \subfigure[ref1][$C_{21}(r_2)$]{\includegraphics[width=.45\linewidth]{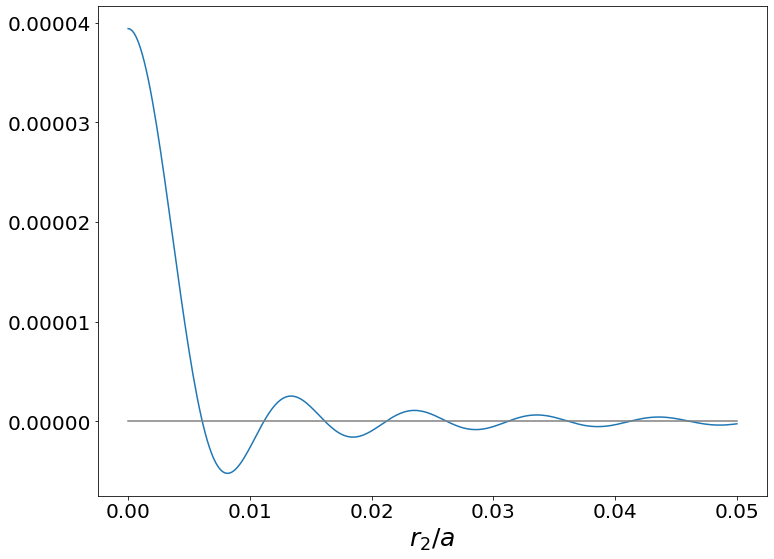}}
	    \caption{Individual $C_{jk}$ factors as a function of $r_2$, using $\partial_T$ to define the normalization. Figures (a) and (b) are the numerical factors of the individual response functions of the first and second detectors, respectively. Figures (c) and (d) are the numerical factors of the crossed response functions $F_{12}$ and $F_{21}$, respectively. Here, $r_1/a = 0$, $|\Delta E|a = 20000$, $a\omega = 1$, $\mu/|\Delta E| = 0.035$ and $\Delta E = -20$ in arbitrary units of energy.}
	    \label{fig:C's}
	\end{figure}

	First, let us study the behavior of the individual terms $C_{11}$ and $C_{22}$ as a function of $r_2$. There is no dependence on $r_2$ on $C_{11}$, so it will be a constant. The $C_{22}$ term is shown in figure \ref{fig:C's}(b), and it has a Bessel dependence on $r_2$, but it is always squared, so it can never be negative. The crossed terms $C_{12}$ and $C_{21}$, on the other hand, has the argument of only one of the Bessel functions varying with $r_2$. As this function has an oscillatory behavior, we also expect the crossed $C$ factors to be oscillatory, as in figures \ref{fig:C's}(c) and \ref{fig:C's}(d). They should have a local (at least) maximum when $r_1 = r_2$ because that's the only point where all the terms in the $m$ and $n$ sums are positive.

\begin{figure}[h!]
	    \centering
	    \subfigure[ref1][$r_1/a = 0$]{\includegraphics[width=.45\linewidth]{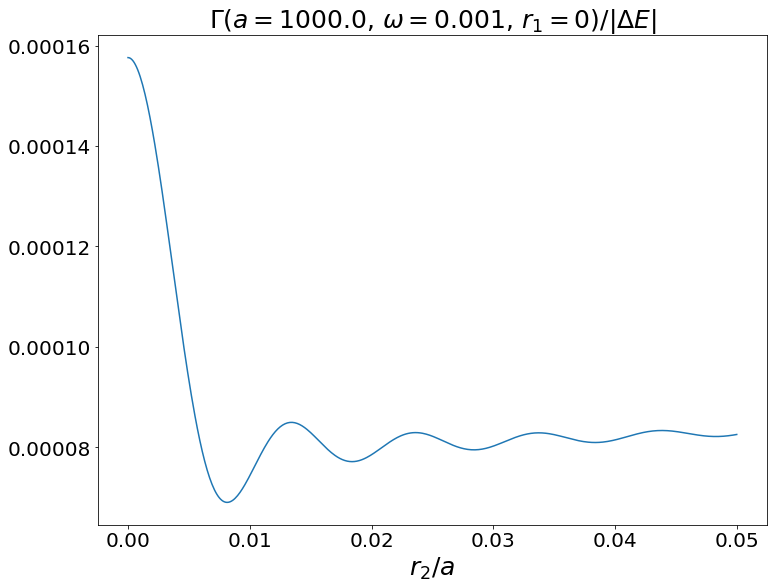}}\qquad
	    \subfigure[ref2][$r_1/a = 0.01$]{\includegraphics[width=.45\linewidth]{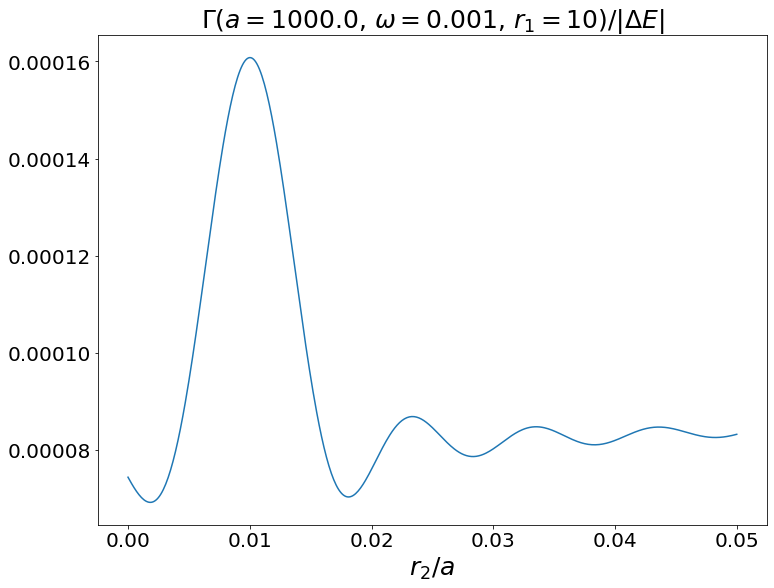}}
	    \caption{Symmetric de-excitation rates for the non-relativistic regime. In the first graph, the first detector is fixed in the origin. In the second graph, the first detector is also fixed, but out of the origin. In both cases, $|\Delta E|a = 20000$, $a\omega = 1$, $\mu/|\Delta E| = 0.035$ and $\Delta E = -20$ in arbitrary units of energy.}
	    \label{fig:r1 in/out symmetric}
	\end{figure}

\begin{figure}[h!]
	    \centering
	    \subfigure[ref1][$r_1/a = 0$]{\includegraphics[width=.45\linewidth]{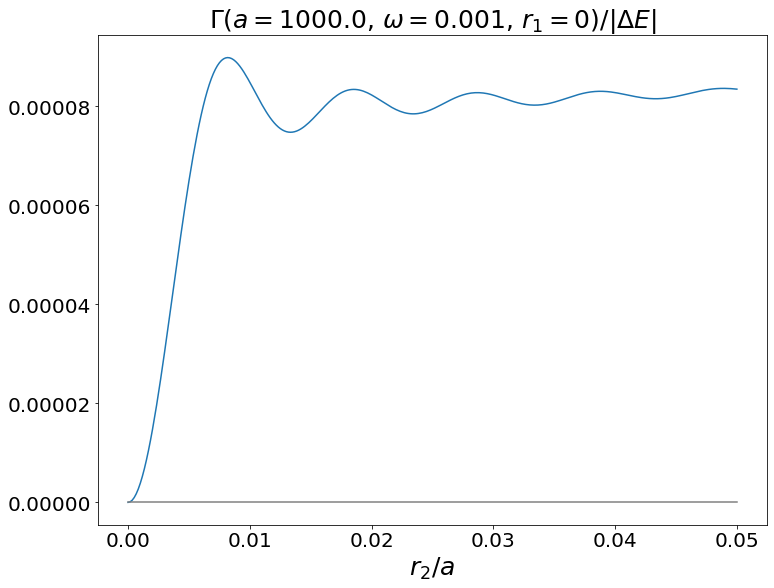}}\qquad
	    \subfigure[ref2][$r_1/a = 0.01$]{\includegraphics[width=.45\linewidth]{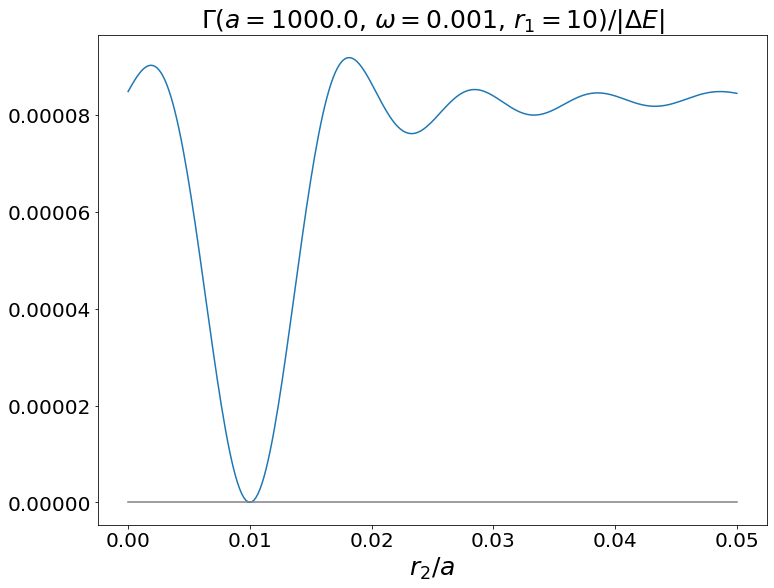}}
	    \caption{Anti-symmetric de-excitation rates for the non-relativistic regime. In the first graph, the first detector is fixed in the origin. In the second graph, the first detector is also fixed, but out of the origin. In both cases, $|\Delta E|a = 20000$, $\omega a = 1$, $\mu/|\Delta E| = 0.035$ and $\Delta E = -20$ in arbitrary units of energy.}
	    \label{fig:r1 in/out anti}
	\end{figure}

	Although its difficult to infer the main properties the transition rate $\Gamma$, in section 4.1 we make an approximation ($\gamma_1 \approx \gamma_2 \approx 1$) to make its behavior more clear and to conclude the existence of a global maximum or a global minimum in the symmetric and anti-symmetric transitions, respectively. This happens because, in the computation of the transition rate, $C_{11}$ and $C_{22}$ are always positive, however the crossed terms $C_{12}$ and $C_{21}$ contributes positively for transitions involving the symmetric state and negatively for transitions involving the anti-symmetric state. In fact, we see in figures \ref{fig:r1 in/out symmetric} and \ref{fig:r1 in/out anti} that there is a global maximum for transitions involving the symmetric entangled state, and a global minimum for transitions involving the anti-symmetric entangled state.

	\begin{figure}[h!]
	    \centering
	    \subfigure[ref1][Non-relativistic regime ($|\Delta E|a = 20000$).]{\includegraphics[width=.47\linewidth]{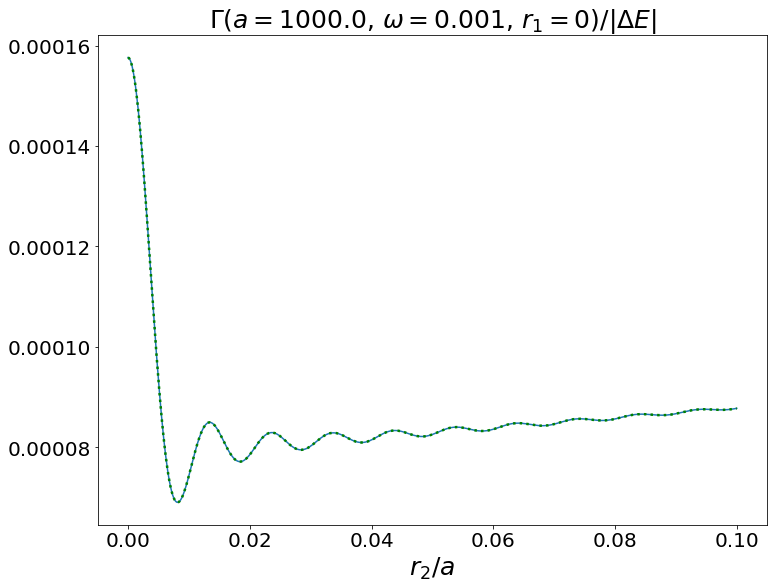}} \qquad
	    \subfigure[ref2][Relativistic regime ($|\Delta E|a = 200$).]{\includegraphics[width=.45\linewidth]{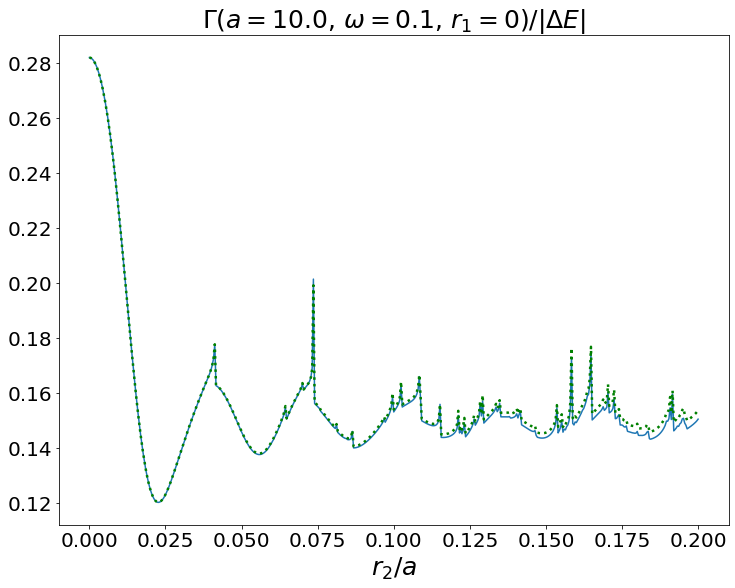}}
	    \caption{Transition rate involving the symmetric entangled state calculated for different normalization constants, as a function of $r_2$. The continuous blue graph was computed using the $\partial_T$ Killing vector, and the dotted green graph was computed using the $\partial_t$ one. The first graph shows the non-relativistic regime, and the second one shows the relativistic regime. In both cases, $r_1/a = 0$, $a\omega = 1$, $\mu/|\Delta E| = 0.035$ and $\Delta E = -20$ in arbitrary units of energy.}
	    \label{fig:CTandCt}
	\end{figure}

	Now, let us calculate the transition rates of the system. First, we need to discuss the normalization used in these calculations. The physical meaning of this choice is the time-like Killing vector used to quantize the massive scalar field, giving the two different normalizations in equation (\ref{eq:normalization}). As discussed in subsection \ref{subsec:discussion}, they only differ by a factor $| 1 - m\omega/\Delta E' |$, which in general is very close to one. In figure \ref{fig:CTandCt} we show the transition rate from $|s\rangle$ to $|g\rangle$ as a function of $r_2$, calculated with both normalizations. We can see that they begin to visually differ for $r_2/a > 0.1$, but there is no qualitative relevant difference. So, in the discussions concerning the dependence on other parameters, we will omit plots using $\partial_t$ as the Killing vector, since they will not provide any further information.

	\begin{figure}[h!]
	    \centering
	    \subfigure[ref1][Non-relativistic regime ($|\Delta E|a = 20000$).]{\includegraphics[width=.47\linewidth]{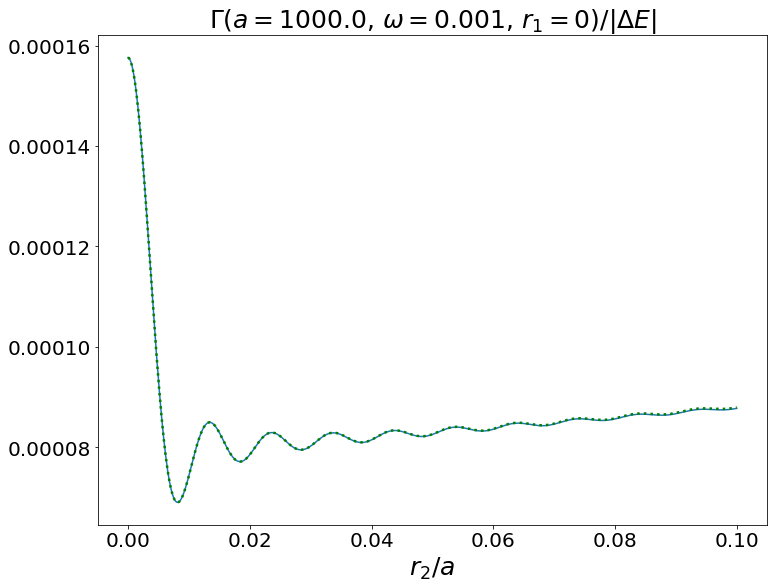}}\qquad
	    \subfigure[ref2][Relativistic regime ($|\Delta E|a = 200$).]{\includegraphics[width=.45\linewidth]{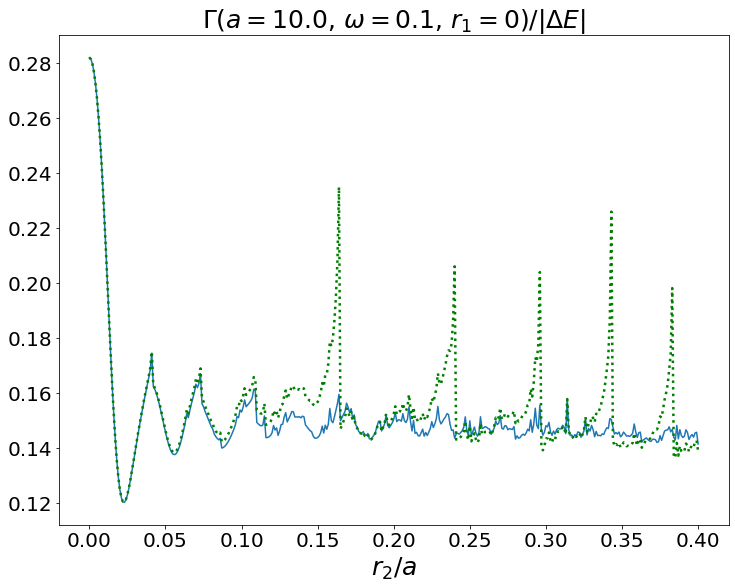}}
	    \caption{Comparison between the transition rates of a de-excitation involving the symmetric entangled state calculated from $C_T$, or using the approximation $\Gamma'_{symm}$. In each graph, the first one is the continuous blue line, and the second one is the dotted green graph. The first graph presents the non-relativistic regime, and the second one presents the relativistic regime. In both cases, $r_1/a = 0$, $a\omega = 1$, $\mu/|\Delta E| = 0.035$ and $\Delta E = -20$ in arbitrary units of energy.}
	    \label{fig:CandS}
	\end{figure}

	Besides that, we can calculate the transitions using the equation (\ref{eq:Ct}) for the $C_T$'s and plugging into the transition rate, or by using directly (\ref{eq:SandA}). In figure \ref{fig:CandS}, we compare a de-excitation involving the symmetric entangled state computed in both ways, with or without the approximation, respectively. In the non-relativistic regime, the graphs are visually identical, but, in the relativistic one, we see that they differ significantly. It was also expected that the peaks were to change, since we also changed the denominator, ignoring a $\gamma$ factor, to get in (\ref{eq:SandA}).

	Let us analyze the different possible transitions and transition rates. We will first study the de-excitations involving the symmetric entangled state. Now, using figure \ref{fig:r1 in/out symmetric}, we compare the rate as a function of $r_2$ in two different situations, when $r_1$ is in or out of the origin, respectively. As expected by the discussion in subsection \ref{subsec:discussion} and by the individual terms in figure \ref{fig:C's}, in both cases we have a maximum when $r_1 = r_2$.

	The behavior of the rate from figure \ref{fig:r1 in/out symmetric}(a) is very similar to other situations studied in the literature \cite{Ford:1994zz}, with the response function oscillating, with a large amplitude only in the first few oscillations. In our problem, the response function does not go to zero when $r_2$ increases. This behavior could be expected since the ``gravitational field" increases at larger distances \cite{fraenkel1979}. Even that the detectors get far away from each other, we still have a growing effect of the ``gravitational field" affecting the system.

	\begin{figure}[h!]
	    \centering
	    \subfigure[ref2][$10^{-4} \leq \omega a \leq 10^{-2}$.]{\includegraphics[width=.45\linewidth]{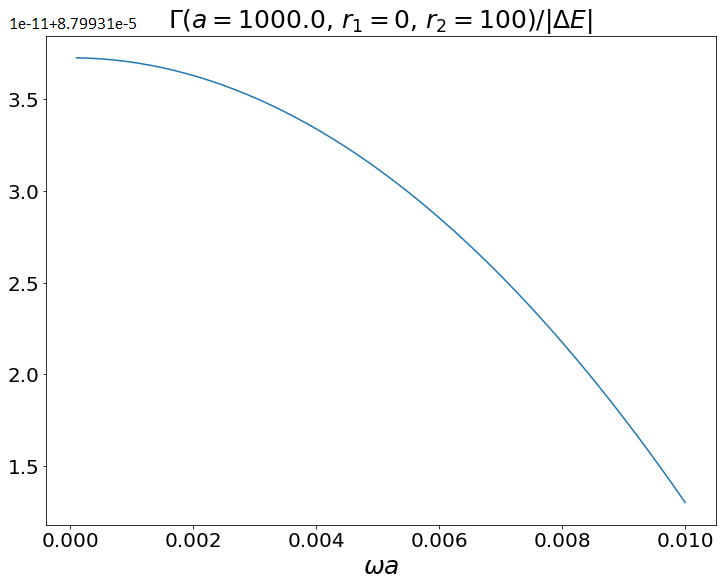}} \qquad
	    \subfigure[ref3][$10^{-2} \leq \omega a \leq 1$.]{\includegraphics[width=.49\linewidth]{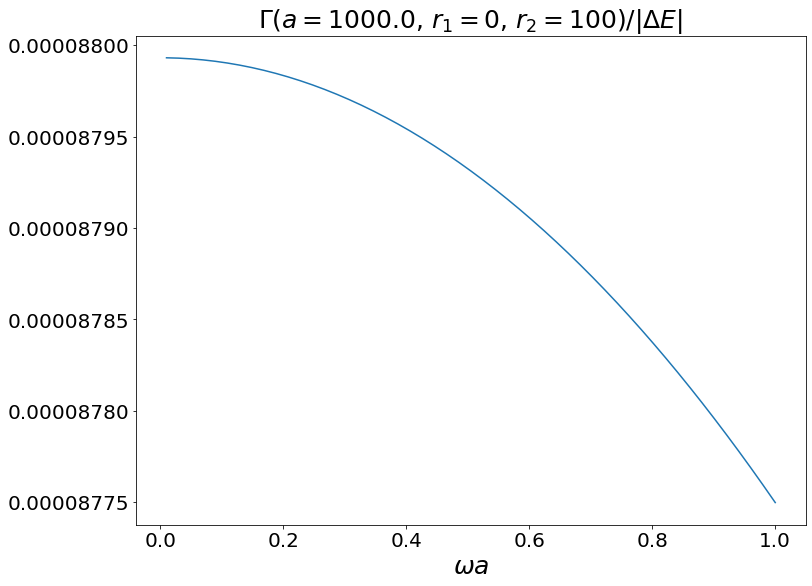}}
	    \caption{Dependence of the transition rate as a function of $\omega$, for $a$ fixed, in the non-relativistic regime ($|\Delta E|a = 20000$). The first graph shows the interval $10^{-4} \leq \omega a \leq 10^{-2}$, and the second graph shows the interval $10^{-2} \leq \omega a \leq 1$. In both cases, $r_1/a = 0$, $r_2/a = 0.1$, $\mu/|\Delta E| = 0.035$ and $\Delta E = -20$ in arbitrary units of energy.}
	    \label{fig:NonRelatSym}
	\end{figure}

	Let us discuss the dependence in $\omega$ in a non-relativistic regime of the system, fixing $|\Delta E|a = 20000$. With other parameters having the same values as in the previous analysis, and now fixing $r_1 = 0$ and $r_2/a = 0.1$, we will take small values of $\omega$, as shown in figure \ref{fig:NonRelatSym}. All the three graphs gives us the same normalized (and very small) transition rate between $8 \times 10^{-5}$ and $9 \times10^{-5}$, coinciding with the value for $r_2/a = 0.1$ in figure \ref{fig:r1 in/out symmetric}(a). Taking $a\omega \leq 10^{-2}$ and $1$, the fluctuations in the rate when $\omega$ changes are respectively 7 and 3 orders of magnitude smaller than the actual value of the rate. So, in this regime, changing $\omega$ basically does not change the rate.

	\begin{figure}[h!]
	    \centering
	    \subfigure[ref2][$10^{-4} \leq \omega a \leq 10^{-2}$.]{\includegraphics[width=.46\linewidth]{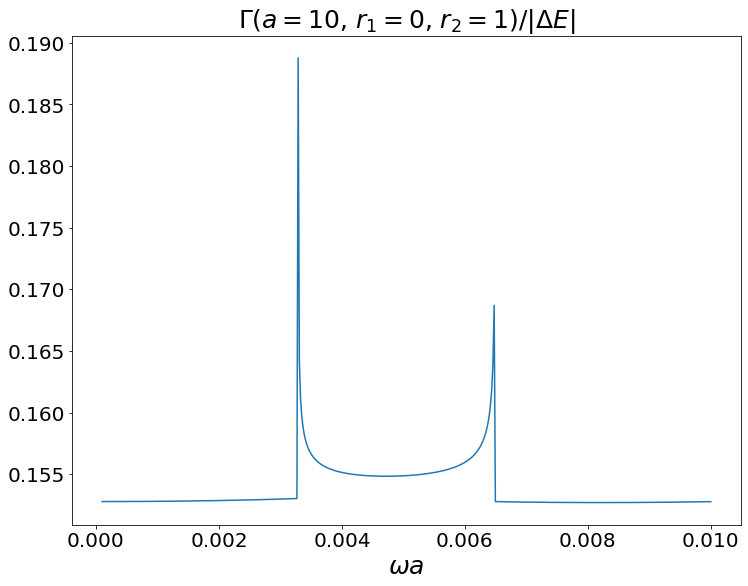}} \qquad
	    \subfigure[ref3][$10^{-2} \leq \omega a \leq 1$.]{\includegraphics[width=.45\linewidth]{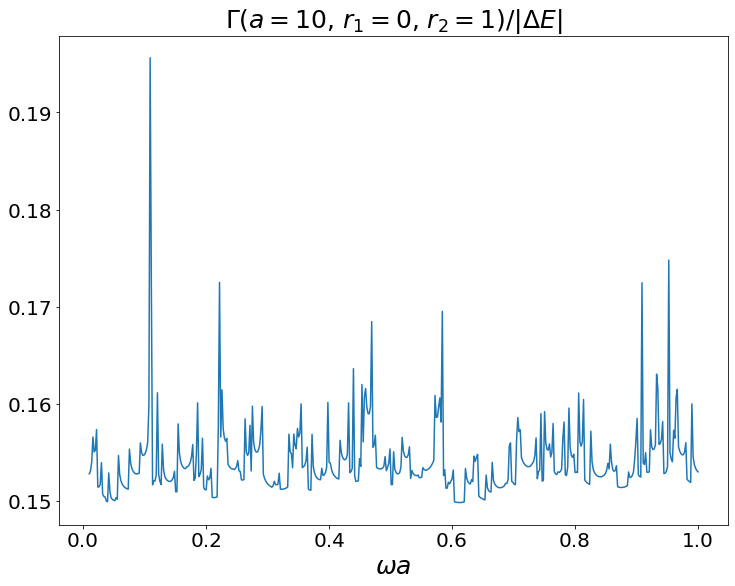}}
	    \caption{Dependence of the transition rate as a function of $\omega$, for $a$ fixed, in the relativistic regime ($|\Delta E|a = 200$). The first graph shows the interval $10^{-4} \leq \omega a \leq 10^{-2}$, and the second graph shows the interval $10^{-2} \leq \omega a \leq 1$. In both cases, $r_1/a = 0$, $r_2/a = 0.1$, $\mu/|\Delta E| = 0.035$ and $\Delta E = -20$ in arbitrary units of energy.}
	    \label{fig:RelatSym}
	\end{figure}

	Now, let us discuss the same dependence in a relativistic regime. Let us fix $|\Delta E|a = 200$, keeping $r_1 = 0$ and $r_2/a = 0.1$. In this case, we can see in figure \ref{fig:RelatSym} that both the rate and its fluctuations are way more relevant than in the non-relativistic one. There are a lot of discontinuities in those graphs as we take $|\Delta E|a = 200$ and $a\omega$ closer to one, since we approach the singularities discussed in subsection \ref{subsec:discussion}, annihilating the denominator $|(m\omega - \Delta E')^2 - k_{mn}^2 - \mu^2|^{1/2}$. But, except for those discontinuities, the dimensionless transition rate does not change significantly with the value of $\omega$ when we fix the other parameters. In the non-relativistic case, with $|\Delta E|a = 20000$, it was roughly $9 \times 10^{-5}$. In the relativistic case, with $|\Delta E|a = 200$, excluding discontinuities, it is always between $0.15$ and $0.16$.

	We can also see, from equations (\ref{eq:CT}) and (\ref{eq:Ct}), that the only significance of the mass of the field, $\mu$, is to change the relativistic regime, changing the zeros of the denominator of the response function.

	\begin{figure}[h!]
	    \centering
	    \subfigure[ref1][Non-Relativistic regime.]{\includegraphics[width=.46\linewidth]{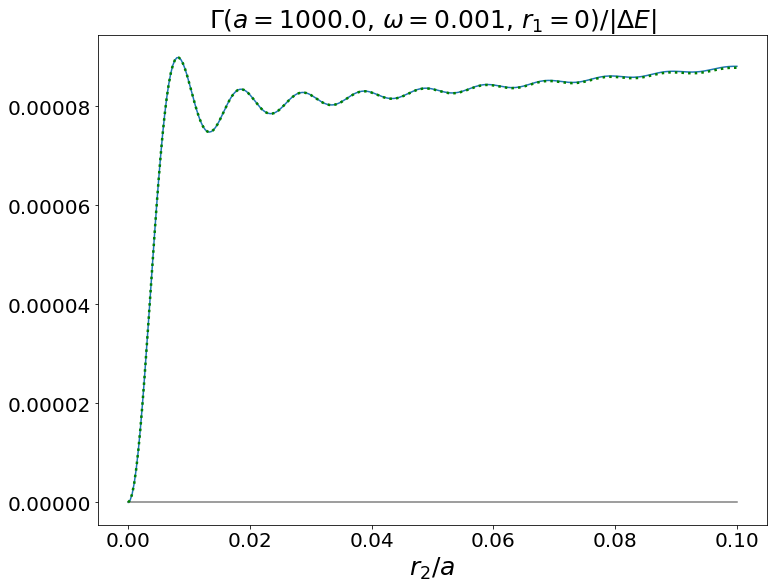}} \qquad
	    \subfigure[ref2][Relativistic regime.]{\includegraphics[width=.45\linewidth]{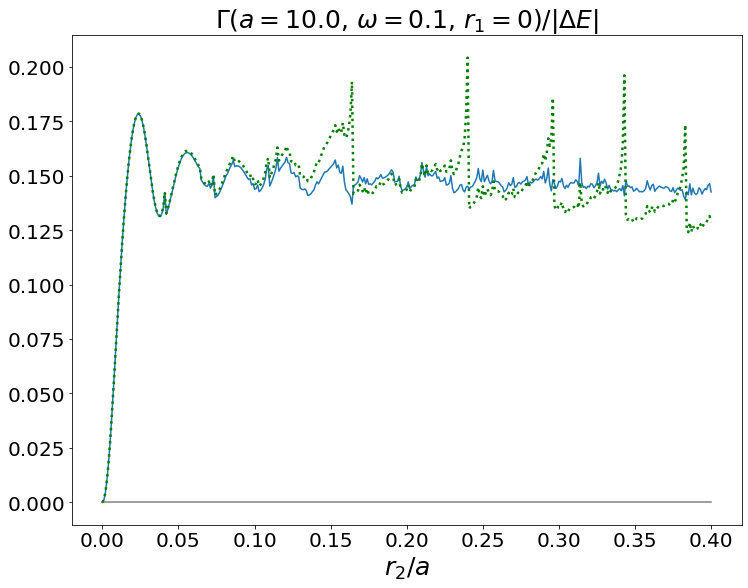}}
	    \caption{Comparison between the transition rates of a de-excitation involving the anti-symmetric entangled state calculated from $C_T$, or using the approximation $\Gamma'_{symm}$. In each graph, the first one is the continuous blue line, and the second one is the dotted green line. The first graph presents the non-relativistic regime ($|\Delta E|a = 20000$), and the second one presents the relativistic regime ($|\Delta E|a = 20$). In both cases, $\omega a = 1$, $\mu/|\Delta E| = 0.035$ and $\Delta E = -20$ in arbitrary units of energy.}
	    \label{fig:CandA}
	\end{figure}

	There is also the de-excitations that involve the anti-symmetric entangled state. In figure \ref{fig:r1 in/out anti}, we show the behavior of the transition rate for this case when the first detector is in the origin or out of the origin, respectively. In figure \ref{fig:CandA}, we show graphs of this transitions computed from the function $C_T$, or from the approximation $\Gamma'_{anti-symm}$. As in the symmetric de-excitation case, the approximation is very good for the non-relativistic regime but very different from the actual transition rate for the relativistic one.

	\begin{figure}[h!]
	    \centering
	    \includegraphics[width=.60\linewidth]{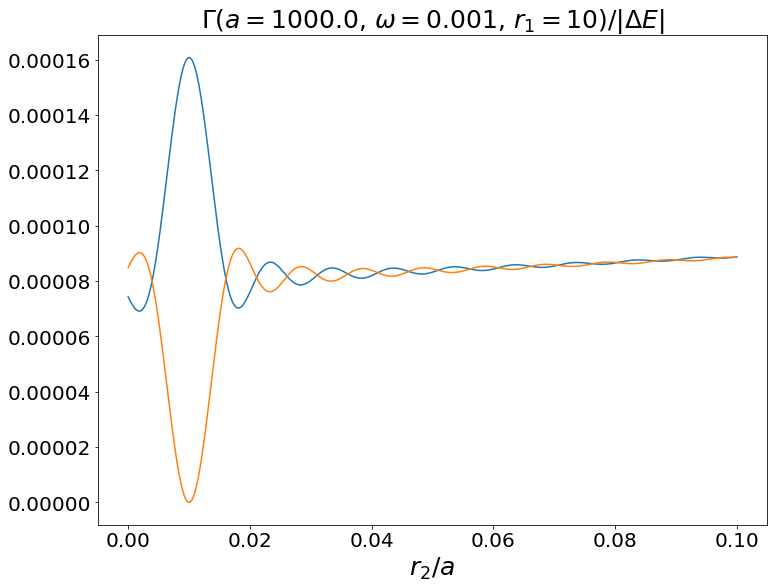}
	    \caption{Comparison between transition rates for de-excitations involving the symmetric and the anti-symmetric states, respectively being the blue continuous line and the orange continuous line, as a function of $r_2$. This is the non-relativistic regime ($|\Delta E|a = 20000$), and the first detector is fixed in $r_1/a = 0.01$. Again, $\omega a = 1$, $\mu/|\Delta E| = 0.035$ and $\Delta E = -20$ in arbitrary units or energy.}
	    \label{fig:Sym/Anti}
	\end{figure}

	Notice that in figure \ref{fig:r1 in/out anti} the anti-symmetric transition rate vanishes for $r_1 = r_2$. In fact, we could think of an intuitive argument for understanding this behavior. The only parameter that distinguishes both detectors in this model is the distance from the origin. If we take equal radii, there are no physical means of distinguishing them. If we interchange both detectors, we do not expect anything to happen to the state. But, in the anti-symmetric entangled state, the system's state should be anti-symmetric if we exchange both detectors. So, it seems not to be possible to have a transition from the excited state to the anti-symmetric entangled state when $r_1 = r_2$.

	Now, for $r_1$ very different from $r_2$, the symmetric and anti-symmetric cases should have very similar transition rates, because  the crossed response functions becomes very small, as seen in figures \ref{fig:C's}(c) and \ref{fig:C's}(d). In figure \ref{fig:Sym/Anti} we explicitly compare the transition rates of de-excitations involving the symmetric and the anti-symmetric entangled states for $r_1/a = 0.01$, and $r_2/a$ varying. Both functions goes to the same value near $8 \times 10^{-5}$ as $r_2 \gg r_1$, with the same behavior.

	If both crossed rates, $R_{12}$ and $R_{21}$, tend to zero, we have only transitions caused by the individual rates. It is expected that even decaying from $|e\rangle$, the final state would not be entangled. In fact, if we have the same transition rate for $|s\rangle$ and $|a\rangle$, with the same sign, we are just generating the pure state $|g\rangle_1 \otimes |e\rangle_2$. But, when $r_1 \approx r_2$, the transitions on the two cases are very different. While the symmetric case displayed a maximum, the anti-symmetric one will display a minimum. In fact, the last one is equal to zero in $r_1 = r_2$, as shown in figure \ref{fig:r1 in/out anti}. Graphs of anti-symmetric transition rates as a function of $\omega$ are visually identical to the graphs in figures \ref{fig:NonRelatSym} and \ref{fig:RelatSym}, so they were omitted.

\begin{figure}[h!]
	    \centering
	    \includegraphics[width=.60\linewidth]{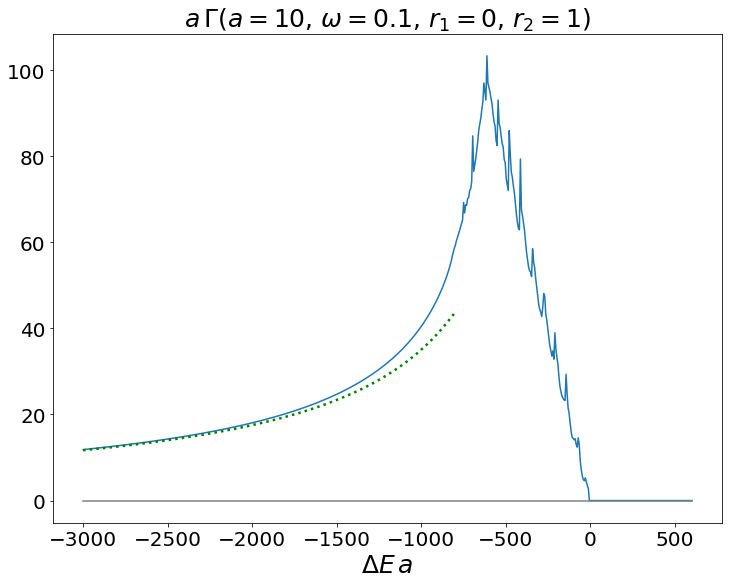}
	    \caption{Dependence of the transition rate on the energy gap of the detector, for $a = 10$ (in arbitrary units of space). The green dotted plot refers to the asymptotic limit $\Delta E \rightarrow -\infty$, where the transition rate goes with $1/|\Delta E'|$. Here, $r_1/a = 0$, $r_2/a = 0.1$, $\omega a = 1$ and $\mu a = 7$.}
	    \label{fig:Energy}
	\end{figure}

	In section \ref{sec:detectors}, we obtained that, for the response function to be different than zero, we needed $\Delta E < 0$. That means we can only see de-excitations in our system. It was already expected, as the rotating vacuum was shown in section \ref{sec:quantization} to be equivalent to the Minkowski vacuum. So, we trivially get that all excitations are identical, and $\Gamma_{|g\rangle \rightarrow |s\rangle} = \Gamma_{|g\rangle \rightarrow |a\rangle} = \Gamma_{|s\rangle \rightarrow |e\rangle} = \Gamma_{|a\rangle \rightarrow |e\rangle} = 0$. But there is a more interesting behavior on the $\Gamma$'s as a function of $\Delta E$. In this discussion, we will take $r_1/a = 0$ different from $r_2/a = 0.1$, so there is no significant difference between transitions involving symmetric or anti-symmetric Bell states, as the crossed rates $C_{12}$ and $C_{21}$ are small compared to $C_{11}$ and $C_{22}$. Let us use a transition involving the symmetric state. We can see in figure \ref{fig:Energy} that there is a gap where transitions are more probable to happen. When the energy of the gap is above some (negative) upper value, there is no transition at all. For negative energy gaps much bigger (in modulus) than $m\omega$, $\alpha_{mn}/a$ and $\mu$, it is expected that the transition rate goes to zero since the rate will be roughly proportional to $1/|\Delta E'|$. Between those limits, it oscillates around a function that steadily grows with the modulus $|\Delta E|$, with a behavior very similar to other works with rotating detectors (see, for example, \cite{Hodgkinson:2014iua}). The extremes of the oscillations depend on the radius $a$ of boundary condition, as it defines the normal modes of the field that mediates the interaction. The asymptotic behavior of the transition rate as $1/|\Delta E'|$ when $\Delta E \rightarrow -\infty$ is illustrated in figure \ref{fig:Energy} by the green dots plotted, as a function $\propto 1/|\Delta E'|$.

	We can try to define the extrema of the interval of $\Delta E$ where the transition rates oscillates by inspection of equation (\ref{eq:CT}). The greater limit of the interval is a specific value defined by the $\Theta$ function, that requires $m\omega - \sqrt{k_{mn}^2 + \mu^2} - \Delta E' > 0$. So, for the other parameters fixed, $\Delta E_{max}$ for having a non-zero transition rate will be given by $\Delta E_{max} = \gamma_2 \cdot \max\limits_{m} (m\omega - \sqrt{k_{m1}^2 + \mu^2})$, already taking the maximum in $n$, for $n = 1$. The minimum value of the interval is related to the regime where $\Gamma$ asymptotically behaves like $1/\Delta E$. That occurs when the term $(m \omega - \Delta E')^2 - k^2_{mn} - \mu^2$ tends to $(\Delta E')^2$, that is, when $\Delta E \ll - \gamma_2 \cdot \max\limits_{m} (m\omega, \alpha_{mn}/a, \mu)$.

	\begin{figure}[h!]
	    \centering
	    \subfigure[ref1][Symmetric Bell state]{\includegraphics[width=.45\linewidth]{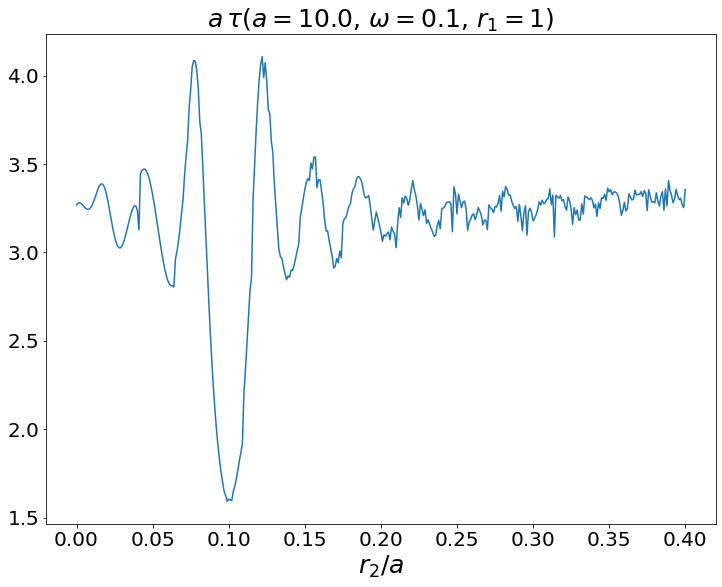}} \qquad
	    \subfigure[ref2][Anti-symmetric Bell state]{\includegraphics[width=.45\linewidth]{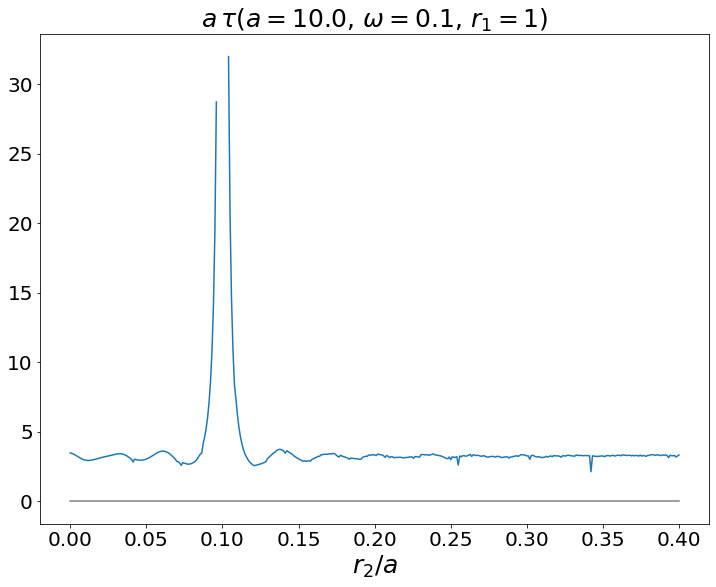}}
	    \caption{The first and second graphs represent the mean-lifes of the symmetric and the anti-symmetric entangled Bell states, respectively, as a function of $r_2$, in the relativistic regime ($|\Delta E|a = 200$), with $\omega a = 1$, $\mu/|\Delta E| = 0.035$ and $\Delta E = -20$ in arbitrary units or energy.}
	    \label{fig:meanlife}
	\end{figure}

	So far we have computed transition rates between states of the system, and we used them to discuss, among other features, the stability of entangled states. A more direct way of analyzing it is to compute the mean life of those states. Since we have no excitation, entangled states can only decay to the ground state $|g\rangle$ of the system, so the mean life of an entangled state $|i\rangle$ is given by

	\begin{equation}
	    \tau_{\, |i\rangle}(r_1, r_2; a, \omega, \Delta E, \mu) = [\Gamma_{|i\rangle \rightarrow |g\rangle}(r_1, r_2; a, \omega, \Delta E, \mu)]^{-1}.
	\end{equation}

	In figure \ref{fig:meanlife}, the behavior of the mean life of both the symmetric and anti-symmetric entangled Bell states is presented. As already pointed out, we see that, for $r_1 = r_2$, the mean life of the symmetric entangled state is a minimum, and the mean life of the anti-symmetric one diverges, as this state becomes stable. For other values of $r_2$, we see the mean life oscillating, with several peaks, as we are in the relativistic regime, but its value is always between $3$ and $3.5$. As $r_2$ gets more different from $r_1$, the amplitudes of the oscillations become smaller. In the non-relativistic regime, the mean life would be a smooth function of $r_2$, since there are no peaks in the transition rate, and consequently no peaks in the mean life.

\section{Conclusions}\label{sec:conclusions} 
	In this work, we studied two entangled Unruh-DeWitt detectors, coupled with a massive scalar field. The radiative processes in a uniformly rotating frame, with Davies-Dray-Manogue's cylinder as the boundary condition for the field, are discussed. Motivated by Davies et al \cite{Davies:1996ks} concerning rates in rotating frames, and by Rodriguez-Camargo et al \cite{Rodriguez-Camargo:2016fbq} -- entanglement of two detectors in a non-inertial frame (Rindler spacetime, in that case) -- we studied radiative processes between two detectors in a rotating frame. Note that the detectors are under the influence of different ``gravitational fields". 

	We extended the Davies et al. result for two entangled detectors, that there can not be any excitation of the detector system in this frame, consistent with the Bogoliubov's $\beta$ coefficients between Minkowski and the rotating field modes being zero. Due to the coupling with the scalar field, there is a non-zero crossed response function. This crossed term is responsible for transitions involving pure states and entangled states of both detectors. We verify that only for $r_1 \approx r_2$, the crossed response functions are significantly different from zero, and this fact was important to study transitions involving entangled states. But first, from the monopole matrices we see that transition rates for de-excitations can be separated into two disjoint cases: the ones involving $|s\rangle$, and the ones involving $|a\rangle$. That is, $\Gamma_{|e\rangle \rightarrow |s\rangle} = \Gamma_{|s\rangle \rightarrow |g\rangle}$ and $\Gamma_{|e\rangle \rightarrow |a\rangle} = \Gamma_{|a\rangle \rightarrow |g\rangle}$. As a consequence of the behavior of the crossed response functions, the second ones tend to zero when $r_2 \rightarrow r_1$, where the first ones have their maximum.

	The entanglement harvesting effect only occurs in the de-excitation $|e\rangle \rightarrow |s\rangle$ between the state where both detectors are excited, and the maximally entangled symmetric state, and only for $r_1 \approx r_2$. In other values of radial coordinates, the crossed response function tends to zero outside this regime, and the transitions to symmetric and anti-symmetric entangled states will have the same rate, generating a statistical pure state.  Entanglement degradation, on the other hand, happens for both the transitions $|s\rangle \rightarrow |g\rangle$ and $|a\rangle \rightarrow |g\rangle$. The only stable state is the anti-symmetric entangled state for $r_1 = r_2$ when the transition rate is equal to zero. We also studied the mean-life of both entangled states. It is also possible to find the stable state by looking at the divergence in the mean-life plot for $r_1 = r_2$, in the anti-symmetric de-excitation. Finally, since there is no excitation, there are no entanglement effects associated with excitations.

	For future works, we have plans to study the radiative processes of two detectors in the scenario of a non-time orthogonal metric, for instance, in the Kerr spacetime, where the effects over radiative processes can be analyzed. From this method, one could analyze the possibility of extracting entanglement from a rotating black hole vacuum. Another possibility is to discuss the degradation of entangled states, and compare with other works about entanglement dynamics in Kerr spacetimes \cite{Menezes:2017oeb}. One could also consider studying these radiative processes with electromagnetic fields and entangled atoms, as more realistic models. Also, more complete treatment for the dynamics of entangled detectors interacting with quantum fields can be given by the master equation approach. These subjects are under investigation by the authors.

	 \acknowledgments This work was partially supported by Conselho Nacional de Desenvolvimento Cient\'{\i}fico e Tecnol\'{o}gico - CNPq, 309982/2018-9 (C.A.D.Z.) and 303436/2015-8 (N.F.S.) and by Funda\c{c}\~{a}o de Amparo \`{a} Pesquisa do Estado do Rio de Janeiro - Faperj, Bolsa Nota 10 201.810/2019 (G.P.).

\bibliography{bibliography}{}

\end{document}